\documentclass[preprint]{acm_proc_article-sp}

\begin{document}
\title{Design, implementation and experiment of a {\ttlit YeSQL} Web Crawler}
\numberofauthors{5}
\author{
\alignauthor Pierre Jourlin\\
       \affaddr{Laboratoire d'Informatique d'Avignon}\\
       \affaddr{Universit{\'e} d'Avignon et des pays de Vaucluse}\\
       \affaddr{BP 1228, 84911 AVIGNON CEDEX, France}\\
       \affaddr{Pierre.Jourlin@univ-avignon.fr}
\alignauthor Romain Deveaud\\
       \affaddr{Laboratoire d'Informatique d'Avignon}\\
       \affaddr{Universit{\'e} d'Avignon et des pays de Vaucluse}\\
       \affaddr{BP 1228, 84911 AVIGNON CEDEX, France}\\
       \affaddr{Romain.Deveaud@univ-avignon.fr}
\alignauthor Eric Sanjuan-Ibekwe\\
       \affaddr{Laboratoire d'Informatique d'Avignon}\\
       \affaddr{Universit{\'e} d'Avignon et des pays de Vaucluse}\\
       \affaddr{BP 1228, 84911 AVIGNON CEDEX, France}\\
       \affaddr{Eric.Sanjuan@univ-avignon.fr}
\and
\alignauthor Jean-Marc Francony\\
       \affaddr{UMR PACTE}\\
       \affaddr{Universit{\'e} Pierre Mend{\`e}s France - Grenoble 2}\\
       \affaddr{Grenoble, France}\\
       \affaddr{jeanmarc.francony@umrpacte.fr}
\alignauthor Fran{\c{c}}oise Papa\\
       \affaddr{UMR PACTE}\\
       \affaddr{Universit{\'e} Pierre Mend{\`e}s France - Grenoble 2}\\
       \affaddr{Grenoble, France}\\
       \affaddr{francoise.papa@umrpacte.fr}
}

\toappear{The copyright of this article remains with the authors.
SIGIR 2012 Workshop on Open Source Information Retrieval.
August 16, 2012, Portland, Oregon, USA.}

\maketitle

\begin{abstract}
We describe a novel, ``focusable'', scalable, distributed web crawler based on GNU/Linux and PostgreSQL that we designed to be easily extendible and which we have released under a GNU public licence. We also report a first use case related to an analysis of Twitter's streams about the french 2012 presidential elections and the URL's it contains.
\end{abstract}




\section{Introduction}
Where scalability is concerned, Apache Nutch\textregistered\footnote{http://nutch.apache.org/} and Heritrix\footnote{https://webarchive.jira.com/browse/HER} are probably the best-known and the most-accomplished open-source web crawlers. They both are sensible choice for Information Retrieval (IR) researchers who intend to build large web corpora. They can be configured to specific needs and can be extended and modified.
However, the Java-language source code\footnote{There are alternatives written in Python, e.g. : Mechanize (36419 lines of code) and Scrapy (23096 lines of code)} of these two software toolkits are rather large and complex: 29349 lines of source 
code for Apache Nutch (v1.4) and 107377 for Heritrix (v3.1.0). Another possible drawback from the researcher's perspective is that they both access the data using unconventional systems : Nutch relies on Hadoop\texttrademark and Heritrix relies on its own code for handling Internet Archive ARC files.

These systems belong to the ``NoSQL'' or ``UnQL'' approaches, supported by the assumption that the widely used SQL relational database standard is a inherent cause of scalability issues. However, this assumption is contested by several database experts. For instance, recent developments around the PostgreSQL project allow it to perform as well as-  and sometimes outperform some - NoSQL databases\cite{Roy:pgcon2010}. This alternative approach has been named YesQL.

By taking profit of the capabilities of a PostgreSQL server, we implemented our web crawler in a total of only 911 lines of C-language code and 200 lines of SQL and PL/pgSQL. At the time this article was written and as far as we know, this is the only available web crawler that is based on PostgreSQL. The tests we performed have shown that instances of the crawler could process over 20 millions of URLs in a few days without beeing noticeably slowed by database operations. We thus believe this web crawler is well worth considering by IR researchers and programmers. 

\section{Software Description}
The source code repository is located at GitHub\footnote{https://github.com/jourlin/WebCrawler} under a GNU public license. 
Everyone can therefore easily download an up-to-date version of the toolkit, provide user's feedback, or join the developer's team. The crawling system can be briefly summarized as follows: 
\begin{itemize}
\item Links and URLs' data are stored in a PostgreSQL\footnote{http://www.postgresql.org/} database. 
\item The user can launch several crawler's instances on several, possibly distant machines. 
\item Each instance of the crawler iteratively:
\begin{enumerate}
\item fetches a list of URLs to be explored by sending a simple SQL query to the database; 
\item downloads the web pages; 
\item extracts new hypertext links to possibly new URLs; 
\item sends the new data back to the server.
\end{enumerate}
\end{itemize}

Figure~1 shows how the communication between internet, web crawler's instances and the PostgreSQL server.
\begin{figure}
\centering
\epsfig{file=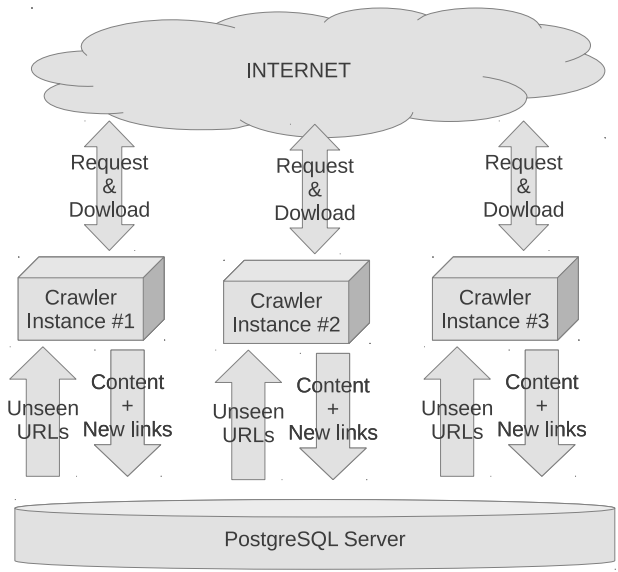, scale=0.4}
\caption{Web crawler organisation}
\end{figure}

The choice of URLs to be fetched is made by one SQL query and two PL/pgSQL additive scoring functions: one scores the URL according to its content, the other scores the URL according to the textual context in which they are linked. The programmer can thus easily implement any \textit{focused} crawling strategy by modifying a single SQL fetch query and two scoring functions. The user can write them in PL/pgSQL in order to take advantage for instance, of PostgreSQL regular expressions. In order to achieve even better performance, he might also write them in C-language and take benefit of PostgreSQL's dynamic loadable objects capability. Figures 2 and 3 show a scoring function in PL/pgSQL that calculates a weighted count of keywords occuring in the URL itself (Figure~2) or in the anchor text that links to it (Figure~3).

\begin{figure}
\centering
\begin{footnotesize}
\begin{verbatim}
CREATE OR REPLACE FUNCTION 
ScoreURL(url url) RETURNS bigint AS 
$$
DECLARE
score INT;
normurl TEXT;
BEGIN
normurl=normalize(CAST(url AS text));
IF CAST(url_top(url) AS TEXT) ='fr' THEN
	score=1;
ELSE
	score=0;
END IF;
IF substring(normurl, 'keyword1') IS NOT NULL THEN
	score=score+2;
END IF;
IF substring(normurl, 'keyword2') IS NOT NULL THEN
	score=score+1;
END IF;
RETURN score;
END;
$$ LANGUAGE plpgsql;
\end{verbatim}
\end{footnotesize}
\caption{A Webcrawler strategy written in PL/PGSQL~: scoring URLs}
\end{figure}

\begin{figure}
\centering
\begin{footnotesize}
\begin{verbatim}
CREATE OR REPLACE FUNCTION 
ScoreLink(context text) RETURNS int AS 
$$
DECLARE
score INT;
normcontext TEXT;
BEGIN
normcontext=normalize(context);
score=0;
IF (substring(normcontext, 'keyword1') IS NOT NULL) THEN
	score = score +1;
END IF;
IF 	(substring(normcontext, 'keyword2') IS NOT NULL) THEN
	score = score +1;
END IF;
RETURN score;
END;
$$ LANGUAGE plpgsql;
\end{verbatim}
\end{footnotesize}
\caption{A Webcrawler strategy written in PL/PGSQL~: scoring links}
\end{figure}

Each crawler instance is only responsible for downloading and processing web pages. The downloading stage is performed by the very mature \emph{GNU/Wget} utility\footnote{http://www.gnu.org/software/wget/}.  The database system is responsible for the coordination of multiple crawlers (thanks to SQL transactions), uniqueness of stored URLs and links (thanks to SQL constraints), crawling strategy (thanks to PL/pgSQL or C functions), etc. Insertions into a single SQL view triggers insertions into the more complex internal table structure.

\section{Use case: coverage of "Tweeted" URLs}
\subsection{Context}
Recent open free network visualisation tools have made easier the qualitative analysis of large social networks\cite{DBLP:conf/icwsm/2010}.
Based on these tools, scientists in humanities can visualize large 
relational data which lead to new hypothesis that will require further network 
crawling and data extraction. 
We show an example of such interaction between 
humanities and computer scientists made possible by our YeSQL crawler. 
  
Political scientists have formulated the hypothesis that for the 2012 French
presidential elections, candidates' communication departments accepted Twitter
as a target media and integrated it to their communication system.

Their strategy was to better control their communication and to improve the 
dissemination of political messages they convey, in order to influence public opinion. 
What was at stake ? The saturation and the meshing of the media sphere, with coherent 
messages whatever the channel of dissemination they choose. 

The empowerment of their communication during the campaign was linked to their capacity~:
\begin{itemize}
\item to consolidate their network of opinion leaders thanks to Twitter, 
\item to be more reactive and to communicate ``just in time'' if unexpected events occur,
\item to strengthen the efficiency of their activists network.
\end{itemize}
As a consequence, the relationships between their different communication devices
has to be analysed. 

\subsection{Experiment}
In order to evaluate this hypothesis, we conducted a capture of Twitter's messages 
and a parallel though independent web crawl of candidate web sites and newspaper's
political pages.
We then attempted to compare the two data sources. Twitter's markers (e.g. '\#' and '@') 
facilitates the production of statistics on a given collection. Regarding the web,
drawing statistics require a very well structured crawl, with good identification
of identical URL and page contents. The YeSQL web crawler proved to be well suited to this 
task.

By filtering tweets from candidates, to candidates or mentioning a candidate (e.g. @fhollande, 
@bayrou, @melanchon2012, @SARKOZY\_ 2012, etc.), we recorded 93592 tweets from february 6th at 00:00am to 
february 13th 2012 at 00:00am. 26638 of those tweets contained a shortened URL (28.4\%) from a set of
10447 unique shortened URL corresponding to 4777 unique effective URLs.

This filtering produced a homogeneous corpus based on a usage logic and identical annunciation rules. 
The reference to candidates' addresses produces a multi-voiced discourse folded up on the proper space 
of Twitter.  Each ``tweeted'' URL  is functioning as an interface with the outside of this space and 
brings back external information from the media space. Their identification is important as a marker 
of discourse evolution and also for its anchorage in the media and political topicality . 

Independently from this collection, we started a web crawler instance that was allowed to download
20 pages in parallel, from february 20th at 00:00am to february 26th at 10:55pm. It was initiated 
on 32 initial URLs from newspapers' political pages and candidates' web sites and collected over 2.7 millions
of URLs. In the following tables, we call "depth" the minimum number of links needed to navigate from an
initial URL (depth=0) to a crawled URL.

Table~1 shows the proportion of tweeted effective URLs that were crawled during this period. The fourth 
column shows that most frequently tweeted URLs are more likely to be covered by the crawl. 
These results show that most popular URLs have a significant probability to be directly retrieved by the crawler
after millions of URLs have been crawled. 

\begin{table}
\begin{center}
\begin{small}
\begin{tabular}{|r|r|r|r|r|}
\hline
Depth	& \# crawled	& \% URLs  &		\% URLs \\
	&	URLs		& covered (a)		& covered (b)\\
\hline
0 &	2 &		0.00 &		0.00\\
1 &	34 &		0.08 &		1.00\\
2 &	1026 &		0.73 &		4.00\\
3 &	8543 &		1.84 &		8.00\\
4 &	56883 &		3.06 &		12.00\\
5 &	368247 &	7.33 &		27.00\\
6 &	2756671 &	15.28 &		40.00\\
\hline
\end{tabular}
\end{small}
\caption{Tweeted URLs' coverage. \small{(a): for all 4777 tweeted URLs ; (b): for the top 100 most frequently tweeted URLs. ``Depth'' is the minimum number of hyperlinks that one has to follow to reach an URL from the initial set.}}
\end{center}
\end{table}

Figure~4 shows the proportion of tweeted URLs found in the crawling per tweeted frequency 
(number of times that the URL was tweeted).
This gives an estimation of the crawling coverage with regard to URL's visibility.
\begin{figure}
\centering
\epsfig{file=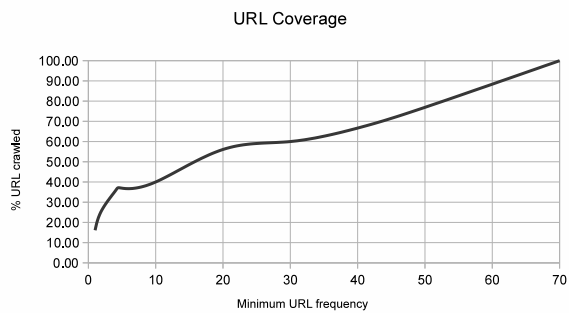, scale=0.5}
\caption{Crawler coverage per tweeted URL frequency.}
\end{figure}

Table~2 shows that the similar problem of tweeted domains instead of tweeted URLs is substantially easier.  
Indeed, the coverage is noticeably higher when only the URL's domains are considered.
In particular, 100 most tweeted domains are almost totally (97.73\%) covered by the web crawl.

More generally, we can observe that high ``domain'' coverage figures are obtained for relatively 
low ``depth'' levels. This suggests that the most popular URLs originates from sites that are the 
nearest neighbours of the 32 initial newspapers' political pages and candidates' web sites.

This is not surprising considering that the web of political blogs is stable along month periods 
\cite{DBLP:conf/icwsm/CointetR10}.
Moreover, all main French newspaper offer a blog service to their readers.
The readers contributions to their websites allow them to capture most of the queries on the web dealing with politics.

\begin{table}
\begin{center}
\begin{small}
\begin{tabular}{|r|r|r|r|r|}
\hline
Depth	& \# crawled	& \% domains &		\% domains \\
	&	domains		& covered (a)		&	covered (b)\\
\hline
0&	1&	0.00&		0.00\\	
1&	31&	2.50&		18.18\\			
2&	95&	4.43&		29.55\\			
3&	312&	11.93&		50.00\\			
4&	1596&	27.73&		81.82\\			
5&	8137&	49.66&		95.45\\			
6&	45992&	72.50&		97.73\\		
\hline
\end{tabular}
\end{small}
\caption{Tweeted URLs' domain coverage. \small{(a): for all 4777 tweeted URLs ; (b): for the top 100 most frequently tweeted URLs. ``Depth'' is the minimum number of hyperlinks that one has to follow to reach an URL from the initial set.}}
\end{center}
\end{table}

Results in Table~2 also allow us to expect much better coverage of URLs by simply launching more crawler's instances, on a single or on multiple machines. 

\section{Conclusion}

The web crawler we presented does not have all the functionalities that offer older and more 
ambitious projects such as Nutch and Heritrix.
However, we have shown that recent functionalities introduced in PostGreSQL about data structures,
triggers and language programming allow to develop powerful web mining tools that can deal
with highly redundant data as well as less frequent signals. 
We illustrated this with a scalable crawler that can explore web networks at a fine grained level.
In particular, this crawler can help in comparing the web to social networks like Twitter. 

In this particular configuration and for this domain, current events about the french electoral campaign
irrigates the two information spaces, the web and Twitter.
The practice of ``tweeting'' URLs becomes usual in the context of modern approaches of information 
reporting and monitoring. 

As we entered this field of investigation by studying the political ``actors'', we saw that a significant 
part of original informations are produced, published and tweeted by these actors. 

We could also question the existence of significant reporting practices outside the control of political apparatus' dissemination strategies. If our results are confirmed in finer grain analysis, we will be able to reconsider the self-organising hypothesis that people tend to associate to social networks.

\bibliographystyle{abbrv}
\bibliography{pgwebcrawler}

\balancecolumns
\end{document}